\documentclass[reprint,superscriptaddress,amsmath,amssymb,prb,longbibliography]{revtex4-1}
\usepackage[scaled=0.92]{helvet}
\usepackage{comment}
\usepackage{color}
\usepackage{graphicx}
\usepackage{subfigure}
\usepackage{dcolumn}
\usepackage{bm}
\usepackage{amsmath}
 
\usepackage{braket}
\usepackage{soul}
\usepackage{diagbox, eqparbox, hhline}
\usepackage[detect-none]{siunitx}
\DeclareMathAlphabet{\mathpzc}{OT1}{pzc}{m}{it}

\renewcommand{\thesection}{\arabic{section}}
\renewcommand{\thesubsection}{\arabic{section}.\arabic{subsection}.}
\renewcommand{\thesubsubsection}{\arabic{section}.\arabic{subsection}.\arabic{subsubsection}}
\usepackage[raggedright]{titlesec}
\usepackage{titlesec}
\usepackage[normalem]{ulem}
\titleformat{\section}
{\color{red}\normalfont\large\bfseries}
{\color{red}\thesection}{0em}{}
\titleformat{\subsection}
{\normalfont\normalsize\bfseries}
{\thesubsection}{0em}{}

\titleformat{\subsubsection}
{\normalfont\normalsize\bfseries}
{\thesubsubsection}{0em}{}
\titlespacing\section{0pt}{10pt plus 4pt minus 2pt}{10pt plus 2pt minus 2pt}
\titlespacing\subsection{0pt}{10pt plus 4pt minus 2pt}{0pt plus 2pt minus 2pt}
\titlespacing\subsubsection{0pt}{10pt plus 4pt minus 2pt}{0pt plus 2pt minus 2pt}

\usepackage{ifluatex,ifxetex}
\ifluatex
  \usepackage{fontspec} 
\else\ifxetex
  \usepackage{fontspec} 
\else 
  \usepackage[T1]{fontenc}
  \usepackage[utf8]{inputenc} 
\fi\fi

\begin{document}

\title{Tailoring High-Frequency Magnonics in Monolayer Chromium Trihalides}

\author{Ra\'i M. Menezes}
\affiliation{Department of Physics, University of Antwerp, Groenenborgerlaan 171, B-2020 Antwerp, Belgium}
\affiliation{NANOlab Center of Excellence, University of Antwerp, Belgium}
\affiliation{Departamento de F\'isica, Universidade Federal de Pernambuco, Cidade Universit\'aria, 50670-901, Recife-PE, Brazil}

\author{Denis \v{S}abani}
\affiliation{Department of Physics, University of Antwerp, Groenenborgerlaan 171, B-2020 Antwerp, Belgium}
\affiliation{NANOlab Center of Excellence, University of Antwerp, Belgium}

\author{Cihan Bacaksiz}
\affiliation{Department of Physics, University of Antwerp, Groenenborgerlaan 171, B-2020 Antwerp, Belgium}
\affiliation{NANOlab Center of Excellence, University of Antwerp, Belgium}

\author{Cl\'ecio C. de Souza Silva}
\affiliation{Departamento de F\'isica, Universidade Federal de Pernambuco, Cidade Universit\'aria, 50670-901, Recife-PE, Brazil}

\author{Milorad V. Milo\v{s}evi\'c}
\email{milorad.milosevic@uantwerpen.be}
\affiliation{Department of Physics, University of Antwerp, Groenenborgerlaan 171, B-2020 Antwerp, Belgium}
\affiliation{NANOlab Center of Excellence, University of Antwerp, Belgium}

\begin{abstract}
Monolayer chromium-trihalides, the archetypal two-dimensional (2D) magnetic materials, are readily suggested as a promising platform for high-frequency magnonics. Here we detail the spin-wave properties of monolayer CrBr$_3$ and CrI$_3$, using spin-dynamics simulations parametrized from the first principles. We reveal that spin-wave dispersion can be tuned in a broad range of frequencies by strain, paving the way towards flexo-magnonic applications. We further show that ever-present halide vacancies in these monolayers host sufficiently strong Dzyaloshinskii–Moriya interaction to scatter spin-waves, which promotes design of spin-wave guides by defect engineering. Finally we discuss the spectra of spin-waves propagating across a moir\'e-periodic modulation of magnetic parameters in a van der Waals heterobilayer, and show that the nanoscale moir\'e periodicities in such samples are ideal for realization of a magnonic crystal in the terahertz frequency range. Recalling the additional tunability of magnetic 2D materials by electronic gating, our results situate these systems among the front-runners for prospective high-frequency magnonic applications.
\end{abstract}

\date{\today}
\pacs{Valid PACS appear here}

\maketitle

\section{. Introduction}

\noindent Two-dimensional (2D) magnetic materials, such as monolayer chromium trihalides and manganese dichalcogenides, have recently drawn immense attention of both theoretical and experimental research, due to their fundamental significance and promising technological applications. The high tunability of the magnetic interactions in such materials, e.g., by lattice straining~\cite{bacaksiz2021distinctive,vishkayi2020strain}, electronic gating~\cite{huang2018electrical,jiang2018controlling}, and layer stacking~\cite{chen2019direct,sivadas2018stacking,shang2019stacking}, among other techniques, directly leads to manipulation of magnetic textures, such as domain-walls, spin-waves (SWs) and magnetic skyrmions, thus opening a range of possibilities for novel device concepts \cite{gong2019two}. Arguably, the largest impact of these manipulations can be realized in the field of magnonics, considering that magnons have been readily detected in atomically-thin chromium tri-iodide~\cite{cenker2021direct}, with documented spin-wave modes even in terahertz frequencies~\cite{jin2018raman}. Further advances on that front would make 2D magnetic materials the prime candidates for ultra-fast information transport and processing based on magnonics. 

\begin{figure}[t!]
\centering
\includegraphics[width=7cm]{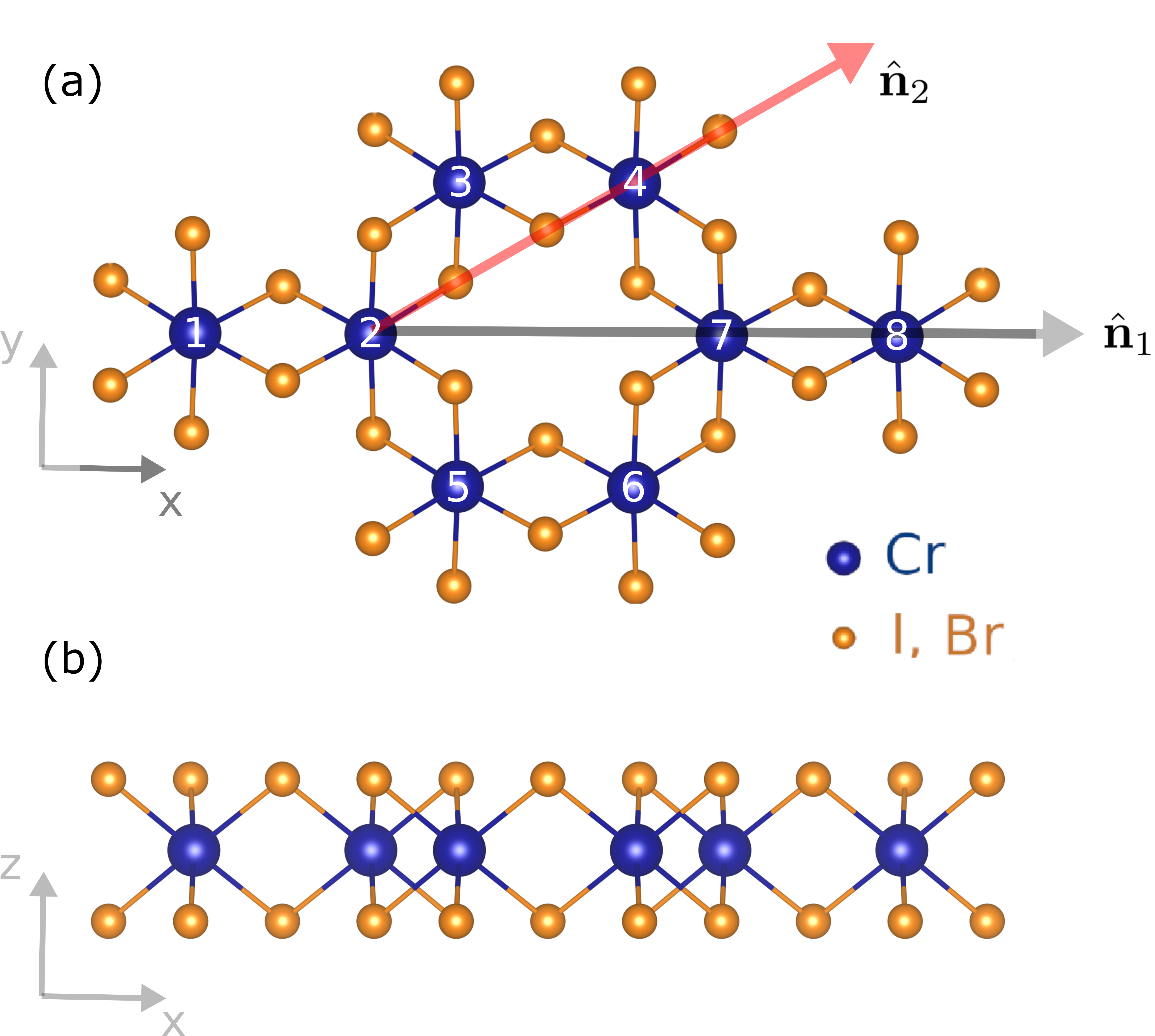}
\caption{(a) Top view of the CrX$_3$ lattice. $\hat{\textbf{n}}_1$ and $\hat{\textbf{n}}_2$ represent the main symmetry axes discussed in Sec.~\ref{sec.dispersion}. (b) Side view of the CrX$_3$ monolayer.  }
\label{figLattice}
\end{figure}

In a chromium trihalide monolayer CrX$_3$ (with X = I, Br or Cl), Cr atoms form a planar honeycomb structure sandwiched between two atomic planes of the halogen atoms, as illustrated in Fig.~\ref{figLattice}. The ferromagnetic super-exchange across the Cr-X-Cr bonds is anisotropic, which together with the weak single-ion anisotropy of Cr results in a ferromagnetic order with off-plane easy axis~\cite{lado2017origin,xu2018interplay}. In addition to the symmetric exchange, the antisymmetric exchange, also known as the Dzyaloshinskii–Moriya interaction (DMI), can be awaken in such systems with any structural or electronic asymmetry seen by Cr atoms, such as present in e.g. Janus monolayers~\cite{xu2020topological} or in applied electric field~\cite{liu2018analysis,behera2019magnetic}.  

In this work we investigate the SW propagation in monolayer CrBr$_3$ and CrI$_3$ as exemplary representatives of 2D magnetic materials. We obtain all relevant magnetic parameters from first-principle calculations on the considered structures~\cite{bacaksiz2021distinctive,sabani2020releasing}, and perform spin-dynamics simulations of the SW propagation. We proceed to calculate the SW dispersion relation in the chromium trihalides under strain, revealing the tunability of magnonic behavior by strain-engineering. We further analyse the role of ever-present lattice defects (halide vacancies) in considered monolayers, usually assumed detrimental for the spin-wave propagation. We show that such lattice defects induce local DMI, which can indeed strongly affect the SW dynamics, but can be proven useful rather than detrimental -- for example, a tailored pattern of defects can confine the SWs and serve as a SW guide. Finally, we report the spectra of SWs propagating across periodic modulation of the magnetic parameters, induced by the moir\'e pattern of a van der Waals heterobilayer~\cite{xiao2021magnetization,tong2018skyrmions}. We show that such samples, readily made experimentally, can function as a magnonic crystal \cite{chumak2017magnonic} for the high-frequency SWs, exhibiting band gaps of propagating SW frequencies. Coupled to the wide range of manipulations available in 2D materials, our results clearly indicate a number of promising directions for terahertz magnonics in magnetic monolayers.

\section{. Atomistic spin simulations}\label{sec.atomistic_model}

\noindent In the simulations we consider a spin system arranged in a honeycomb structure in order to mimic the magnetic moments of Cr atoms in the chromium trihalide monolayer. To reduce the magnetic behavior of the entire monolayer to just the simulation within the Cr plane is made possible by \textit{ab initio} parametrization of microscopic magnetic interactions for different considered cases, done according to Refs.~\onlinecite{bacaksiz2021distinctive,sabani2020releasing}, where we performed \textit{ab initio} four-state energy mapping (4SM) based on density functional theory (DFT) calculations. For more details on the method we refer to Ref.~\onlinecite{vsabani2020ab}. 

In the spin simulations we consider the quadratic Heisenberg spin Hamiltonian, given by
\begin{equation}
    \mathcal{H}=\frac{1}{2}\sum_{i,j}\textbf{S}_i\mathcal{J}_{ij}\textbf{S}_j +\sum_i\textbf{S}_i\mathcal{A}_i\textbf{S}_i.
    \label{eq.H}
\end{equation}
Here $\mathcal{J}_{ij}$ and $\mathcal{A}_i$ are the exchange and single ion anisotropy (SIA) matrices, respectively, and $\textbf{S}_i=(S_i^x,S_i^y,S_i^z)$ is the spin vector at the $i^{\rm th}$ site. We consider $S = 3/2$ for the Cr$^{3+}$ ions, with $3$ unpaired valence electrons and quenched orbital moment ($L=0$), which yields a magnetic moment of $\sim3\mu_B$ per Cr atom, in agreement with the experimental observations\cite{mcguire2015coupling,kim2019micromagnetometry,lado2017origin,xu2018interplay,vsabani2020ab}.

The sum over $i$ in Eq. \eqref{eq.H} runs over all Cr sites, while the sums over $i, j$ run over all nearest-neighbor Cr pairs. The exchange matrix can be further decomposed into a symmetric exchange $J$ and the antisymmetric DMI vector $\textbf{D}$. The Hamiltonian then becomes
\begin{equation}
\begin{aligned}
    \mathcal{H}=&\frac{1}{2}\sum_{i,j}\left[J_\alpha S_i^\alpha S_j^\alpha + J_\beta S_i^\beta S_j^\beta + J_\gamma S_i^\gamma S_j^\gamma\right.\\
    &+ \textbf{D}_{ij}\cdot(\textbf{S}_i\times\textbf{S}_j) \Big]\\
    &+\sum_i\left[\mathcal{A}_{\alpha'}(S_i^{\alpha'})^2+\mathcal{A}_{\beta'}(S_i^{\beta'})^2+\mathcal{A}_{\gamma'}(S_i^{\gamma'})^2 \right],
    \label{eq.H2}
\end{aligned}
\end{equation}
where \{$\alpha\beta\gamma$\} and \{$\alpha'\beta'\gamma'$\} are the local bases of eigenvectors that diagonalize $J$ and $\mathcal{A}$, respectively. $S_i^u=\frac{\textbf{u}\cdot\textbf{S}_i}{\textbf{u}\cdot\textbf{u}}$ is the projection of the $i^{th}$ spin along the vector $\textbf{u}$. For simplicity, and since in this work we are interested in high frequency SWs dominated by the short-range exchange interactions, we neglect the contributions of dipole-dipole interactions. Notice that the above Hamiltonian is not limited to isotropic exchange interactions, and can be applied to different atomistic structures, e.g. including lattice defects where the bases of eigenvectors can change for specific bonds. For either considered structure, corresponding exchange matrices for different Cr pairs are obtained from first principles calculations. Table~\ref{table_parameters} lists the magnetic parameters obtained for pristine CrI$_3$ and CrBr$_3$ monolayers. Parameters obtained for other specific cases considered in this paper will be shown where needed.

To simulate the magnonic behavior, we resort to spin dynamics captured by the Landau-Lifshitz-Gilbert (LLG) equation
\begin{equation}
    \frac{\partial \hat{\textbf{S}}_i}{\partial t}= -\frac{\gamma}{(1+\alpha^2)\mu}\left[\hat{\textbf{S}}_i\times\textbf{B}_i^\text{eff}+\alpha\hat{\textbf{S}}_i\times(\hat{\textbf{S}}_i\times\textbf{B}_i^\text{eff}) \right],
    \label{eq.LLG}
\end{equation}
where $\gamma$ is the electron gyromagnetic ratio, $\alpha$ is the damping parameter, $\mu$ is the magnetic moment per Cr atom, and $\textbf{B}_i^\text{eff}=-\partial \mathcal{H}/\partial\hat{\textbf{S}}_i$ is the effective field. In this work, the LLG spin dynamics simulations are primarily based on the simulation package \textit{Spirit} \cite{muller2019spirit}, adapted to accommodate our (Kitaev) Hamiltonian [Eq.~\eqref{eq.H2}].

\begin{table}[t!]
\begin{tabular}{lcccccccccc}
\hline\hline
  &  pair     &$J^{xx}$& $J^{yy}$ & $J^{zz}$ & $J^{xy}$ & $J^{xz}$  & $J^{yz}$  \\
  &  ($i$-$j$)& (meV) & (meV)   & (meV)  & (meV)              &    (meV)         &    (meV)           \\
\hline
CrI$_3$&                   &       &         &        &                     &                   &        \\
       & (1-2)  & -5.10 &  -3.72  &  -4.63 &   0.00              & 0.00              & 0.84         \\
       & (2-3)  & -4.07 &  -4.76  &  -4.63 &  -0.60              & 0.72              &-0.42         \\
       & (2-5)  & -4.07 &  -4.76  &  -4.63 &   0.60              &-0.72              &-0.42        \\
&$\langle J \rangle$     & -4.41 &  -4.41  &  -4.63 &   0.00              & 0.00              & 0.00        \\
CrBr$_3$&                   &       &         &        &                     &                   &  \\
       & (1-2) & -3.45 &  -3.29  &  -3.42 &   0.00              & 0.00              & 0.10         \\
       & (2-3)  & -3.33 &  -3.41  &  -3.42 &  -0.07              & 0.08              &-0.05         \\
       & (2-5)  & -3.33 &  -3.41  &  -3.42 &   0.07              &-0.08              &-0.05         \\
&$\langle J \rangle$& -3.37 &  -3.37  &  -3.42 &   0.00              & 0.00              & 0.00        \\
\hline\hline
       & $\Delta$ & $\mathcal{A}^{zz}$  &         &        &                     &                   &         \\
       & (meV) & (meV)  &         &        &                     &                   &          \\
\hline
CrI$_3$&  -0.22&    -0.07    &         &        &                     &                   &        \\
CrBr$_3$&  -0.04 &  -0.01    &         &        &                     &                   &         \\
\hline\hline 
\end{tabular}
\caption{Magnetic parameters for pristine CrI$_3$ and CrBr$_3$ obtained from first-principle calculations [see Ref.~\onlinecite{bacaksiz2021distinctive} for details]. $J^{xx}$, $J^{yy}$, and  $J^{zz}$ are diagonal elements, and $J^{xy} = J^{yx}$, $J^{xz} = J^{zx}$, $J^{yz} = J^{zy}$ the off-diagonal elements of the exchange matrix. $\langle J \rangle$ is the average exchange over the three nearest-neighbour pairs and $\Delta=\langle J^{xx}\rangle - \langle J^{zz}\rangle$ is the out-of-plane anisotropy. $\mathcal{A}^{zz}$ is SIA parameter, same for each Cr site.  Pairs (\textit{i-j}) are indicated in Fig.~\ref{figLattice}~(a).}
\label{table_parameters}
\end{table}

\section{. Strain-tunable spin-wave propagation in CrX$_3$ monolayers}\label{sec.dispersion}

\begin{figure}[t!]
\centering 
\includegraphics[width=\linewidth]{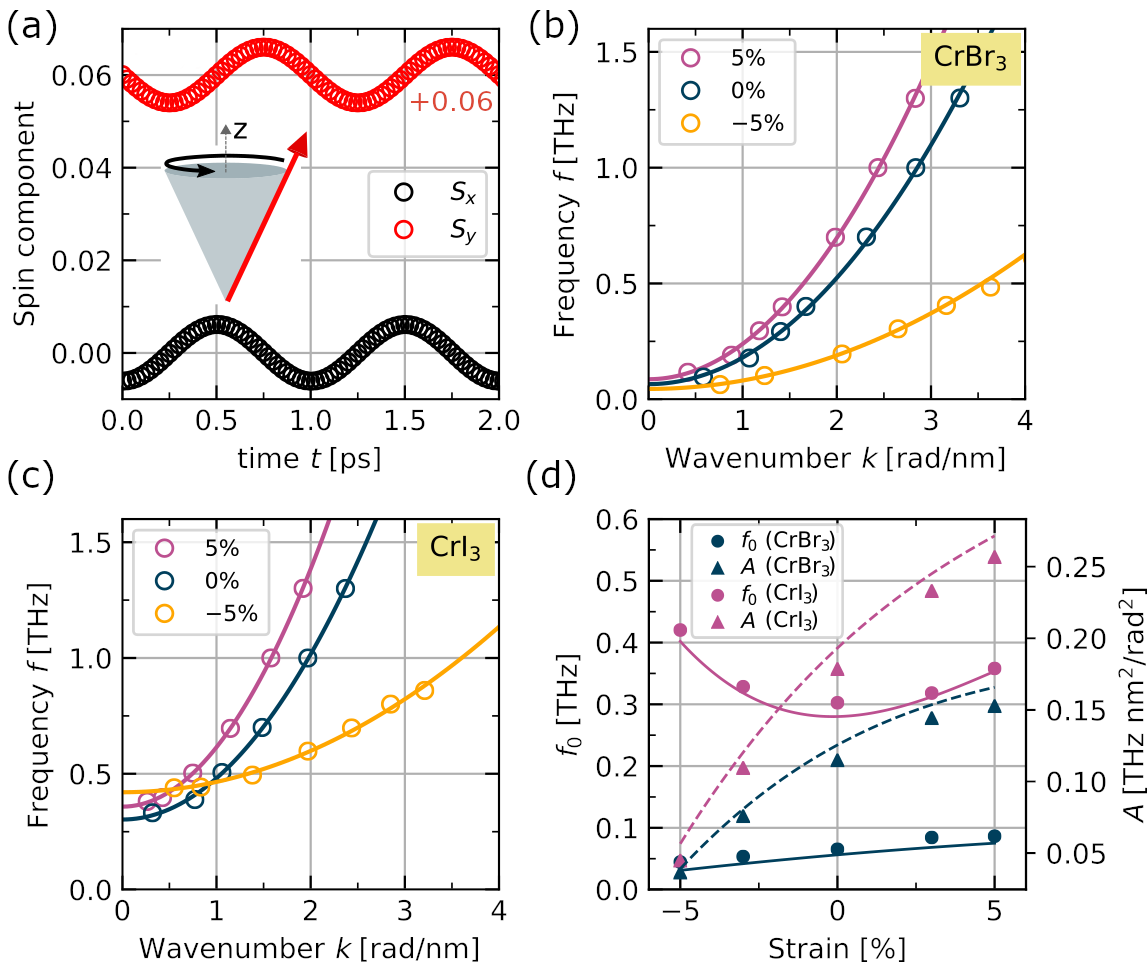}
\caption{(a) Temporal evolution of the spin components for the case of a $1$~THz spin wave. The $S_y$ spin component is vertically shifted for better visualization. (b,c) Spin-wave dispersion relation obtained in the simulations for CrBr$_3$ (b) and CrI$_3$ (c), under different strain. Solid lines show the numerical fit to the quadratic expression $f(k)=A k^2+f_0$. (d) The values of the fitted parameters $A$ and $f_0$ obtained as a function of strain. Solid and dashed lines are the analytical expressions (obtained from Eq.~(\ref{eq.w})) for $f_0$ and $A$, respectively. }
\label{fig2}
\end{figure}

\noindent The magnetic interactions in 2D materials are very sensitive to any deformation in the atomic lattice. In particular, it has been shown that exchange interactions in CrI$_3$ and CrBr$_3$ are significantly affected by either tensile or compressive strain~\cite{bacaksiz2021distinctive}. Therefore, in this section, we discuss how straining the material can affect the propagation of SWs in magnetic monolayers. For the simulations, we consider both CrBr$_3$ and CrI$_3$ monolayers in their pristine form, as well as under uniform biaxial strain. The SW beams are artificially created by a sinusoidal in-plane oscillating field $\textbf{B}_\text{input}=b_0\sin(2\pi f_\text{in} t)\hat{k}$ applied in a narrow rectangular region~\cite{gruszecki2016microwave,song2019omnidirectional}, where $f_\text{in}$ is the input frequency, $b_0$ the field amplitude, and $\hat{k}$ the SW propagation direction. In the simulations we consider $b_0=0.1$~T and damping parameter $\alpha=0.001$. The SW frequency $f$ and wavelength $\lambda$ are calculated by fitting a sine function to the magnetization oscillations [see e.g. Fig.\ref{fig2}~(a)] as a function of time and space respectively. Fig.~\ref{fig2}~(b,c) shows the dispersion relation, i.e., the relation between the SW frequency and the wavenumber $k=2\pi/\lambda$ for CrBr$_3$ and CrI$_3$, respectively, obtained for materials under different strain. The calculated dispersion curves correspond to the lower-energy magnon modes. Notice that the SW dispersion can be tuned throughout a wide range of frequencies by straining the honeycomb structure, which actually demonstrates the ease of tuning magnonics in such 2D materials. Solid lines in Fig.~\ref{fig2}~(b,c) depict the fit of the quadratic expression $f(k)=A k^2+f_0$ to the numerical data. The values of the fitting parameters $A$ and $f_0$ as a function of strain are shown in Fig.~\ref{fig2}~(d). One can clearly notice the distinctly different response of the parameter $f_0$, which corresponds to the zero-momentum SW mode, to strain in CrBr$_3$ and CrI$_3$. As discussed below, this behaviour is directly related to the out-of-plane exchange anisotropy, whose magnitude for CrI$_3$ increases for either tensile or compressive strain, while for CrBr$_3$ it changes linearly with strain [see, e.g., Fig S1 in the supplemental material~\cite{SM}]. The corresponding energy gap at $k=0$, given by $\Delta E=hf_0$, where $h$ is the Plank constant, varies in the range of $1.25$ to $1.74$~meV for CrI$_3$ and from $0.18$ to $0.36$~meV for CrBr$_3$, under the $-5$ to $5$~\% range of strain. Those values have the same order of magnitude, but are significantly smaller than those measured for the case of few layer CrI$_3$ in Ref.~\onlinecite{jin2018raman} ($9.4$~meV) and Ref.~\onlinecite{klein2018probing} ($3$~meV). Despite this quantitative discrepancy, we expect that the here reported strain engineering of spin-wave excitations, manifesting differently in different Cr-trihalide monolayers, can be validated by experimental techniques~\cite{chen2017group,qin2019experimental,klein2018probing,jin2018raman,chen2018topological}.

\begin{figure*}[t!]
\centering
\includegraphics[width=\linewidth]{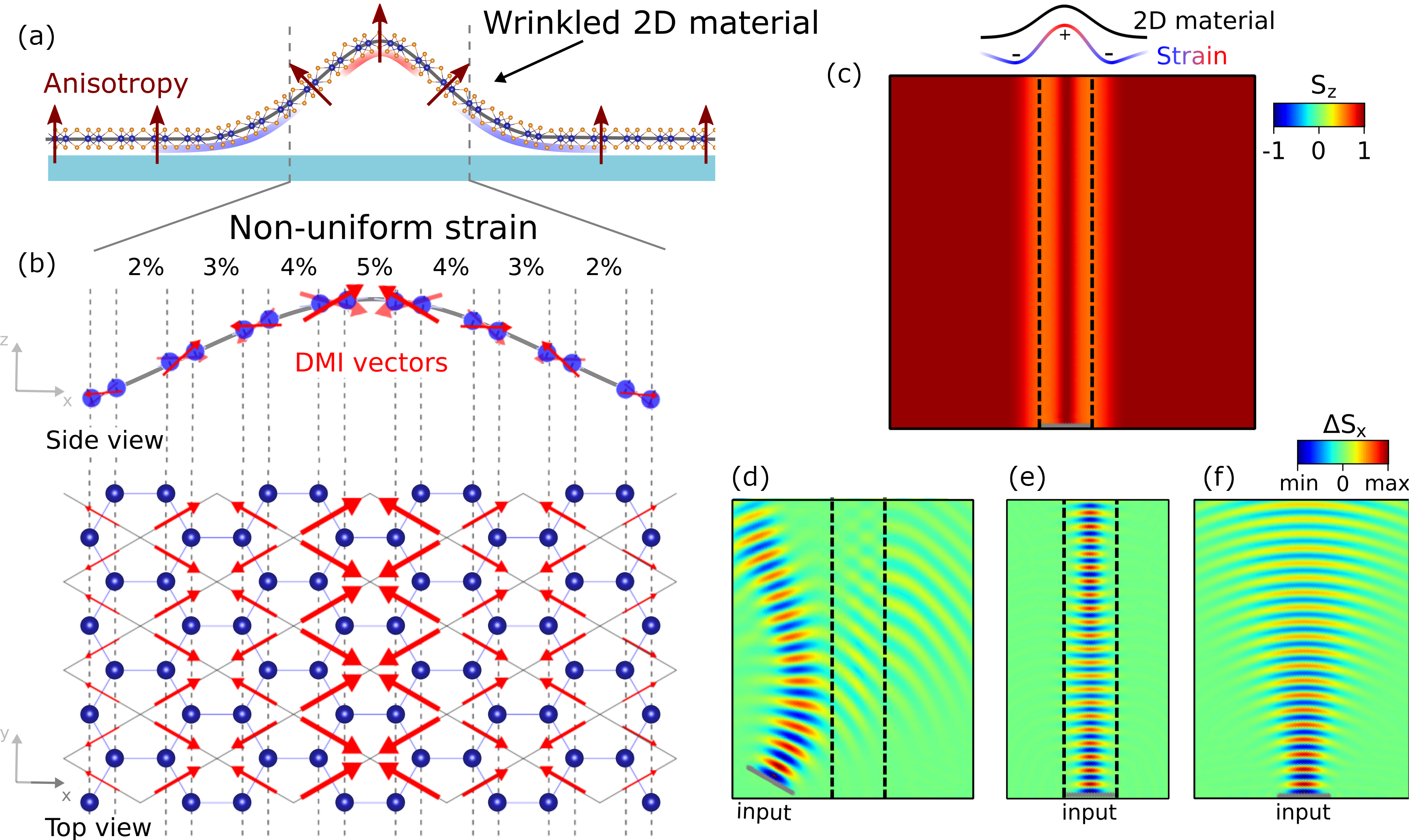}
\caption{ (a) Schematic view of a wrinkled 2D magnetic material. The induced curvature is nonuniformly straining the 2D material, while the magnetic anisotropy remains orthogonal to the monolayer surface. Here, the red shade shows tensile strain and the blue shade represents the compressive strain. (b) Illustration of the DMI pattern expected in the case of nonuniform strain in the CrX$_3$ monolayer \cite{sabani2020releasing}. (c) Snapshot of the simulated magnetic ground state around a wrinkle in the 2D material. Dashed lines indicate the maximal vertical slope in the 2D material. (d-e) Snapshots of simulated SW propagation (d) across and (e) along a wrinkle in CrI$_3$, with $f=0.5$~THz. (f) Snapshot of the SW simulation corresponding to (e), but for pristine (flat) monolayer. }
\label{fig_wrinkle}
\end{figure*}

The SW dispersion can be obtained analytically in the limit of small SW amplitude by solving the linearized LLG equation~\cite{rezende2020fundamentals}. For that purpose, let us simplify the Hamiltonian [Eq.~(\ref{eq.H2})] as follows. First notice that, due to the symmetry of the lattice, the DMI contribution is annulled, and the only nonzero term of SIA matrix is $\mathcal{A}^{zz}$,~\cite{vsabani2020ab} such that the SIA contribution to the energy of the $i^{th}$ spin becomes $\mathcal{A}^{zz}(S_i^z)^2$. Assuming that both CrBr$_3$ and CrI$_3$ have strong out-of-plane anisotropy, the exchange energy can be rewritten in terms of an isotropic exchange $J_0=\langle J^{xx}\rangle$, and the anisotropic term $\Delta=\langle J^{zz}\rangle-\langle J^{xx}\rangle$, where $\langle...\rangle$ represents the average over the three nearest-neighbour pairs. The lattice symmetry guarantees that $\langle J^{yy}\rangle=\langle J^{xx}\rangle=J_0$ and $\langle J^{ab}\rangle=0$ for $a\neq b$ [see, e.g., Table~\ref{table_parameters}]. The Hamiltonian for the $i^{th}$ spin takes the simple form of the XXZ model~\cite{zhang2021two,torelli2019high,lado2017origin}
\begin{equation}
    \mathcal{H}_i=J_0\sum_j\textbf{S}_i\cdot\textbf{S}_j + \Delta\sum_jS_i^z S_j^z + \mathcal{A}^{zz}(S_i^z)^2,
    \label{eq.Hi}
\end{equation}
where the sum in $j$ runs over the three nearest-neighbours of the $i^{th}$ spin. The SW dispersion is then calculated by assuming the linearized solution $\hat{S}^z\approx 1$,  $\hat{S}^x\approx A_0e^{i(\textbf{k}\cdot\textbf{r}-\omega t)}$ and $\hat{S}^y\approx iA_0e^{i(\textbf{k}\cdot\textbf{r}-\omega t)}$, where $A_0\ll1$ represents the SW amplitude; $\omega$ is the SW angular frequency and $\textbf{r}$ is the position of the considered spin. Substituting that into the LLG equation [Eq.~(\ref{eq.LLG})], with the effective field derived from Eq.~(\ref{eq.Hi}), we obtain~\cite{SM}
\begin{equation}
   \omega(k) = \frac{\gamma S^2}{(1+\alpha^2)\mu}\left[ 3\Delta +2\mathcal{A}^{zz}+J_0 g(\textbf{k}) \right],
    \label{eq.w}
\end{equation}
with
\begin{equation}
    \begin{aligned}
        g(\textbf{k})=&3-\cos(ka)-2\cos(ka/2),\quad \text{if }\textbf{k}\parallel\hat{\textbf{n}}_1,\\
        g(\textbf{k})=&2-2\cos(ka\sqrt{3}/2),\quad \text{if }\textbf{k}\parallel\hat{\textbf{n}}_2,
    \end{aligned}
    \label{eq.gk}
\end{equation}     
where $a$ is the distance between nearest-neighbour pairs, and $\hat{\textbf{n}}_1$ and $\hat{\textbf{n}}_2$ are the main symmetry axis of the honeycomb lattice, as shown in Fig.~\ref{figLattice}. 

Notice that in the limit of small $ka$ Eq.~(\ref{eq.gk}) can be approximated as $g(\textbf{k})\approx\frac{3a^2}{4}k^2$ for both $\textbf{k}\parallel\hat{\textbf{n}}_1$ and $\textbf{k}\parallel\hat{\textbf{n}}_2$. The SW frequency $f=\omega/2\pi$ then assumes quadratic dependence on the wavenumber $k$. The solid and dashed lines in Fig.~\ref{fig2}(d) show the analytical solutions (with $\alpha^2\ll1$) for the zero-momentum SW frequency $f_0=\frac{\gamma S^2}{2\pi\mu}(3\Delta+2\mathcal{A}^{zz})$ and the quadratic coefficient $A=\frac{\gamma S^2}{2\pi\mu}\frac{3a^2J_0}{4}$, respectively, which are in very good agreement with the numbers obtained in the simulations. The values of $\Delta$, $\mathcal{A}^{zz}$ and $J_0$ as a function of strain are shown in Fig. S1 of the supplemental material~\cite{SM}. One thus concludes that the XXZ model [Eq.~(\ref{eq.Hi})] is a suitable approximation to describe properties of SWs in uniform CrI$_3$ and CrBr$_3$ 2D magnets. 

In the case where the CrX$_3$ lattice experiences non-zero DMI, such as in Janus structures~\cite{xu2020topological} or in the presence of out-of-plane applied electric field~\cite{liu2018analysis,behera2019magnetic},
the linearized solution for the SW dispersion [Eq.~(\ref{eq.w})] results in an extra term that has linear dependence on the wavevector $k$.~\cite{cortes2013influence} The contribution of DMI to the SW dispersion depends on the angle between the DMI vector and the axis around which the spins rotate, and the strongest effect is observed when the magnetization lies in the same plane as the DMI vectors. Details on the calculation of SW dispersion in presence of uniform DMI are provided in the supplemental material~\cite{SM}.

\section{. Flexo-magnonics}\label{sec.wrinkle}

The strong response of SW properties to strain in the 2D material suggests the possibility of flexo-magnonic applications. In fact, 2D materials are mostly very flexible and can be strain engineered in a multitude of ways. For example, localized strain can be induced by growing the 2D material on top of a patterned substrate~\cite{wang2019strain,peng2020strain}, or by placing it onto elastometric or piezoelectric substrate, whose compression may lead to wrinkling and buckling of the 2D material~\cite{castellanos2013local,sun2019strain,peng2020strain}. In addition, bubbles and tents can be formed in the 2D material by trapping water, gas or solid nanoparticles at the interface between the magnetic monolayer and the substrate~\cite{khestanova2016universal,dai2019strain}. 

To exemplify the interaction of SWs with such localized heterostrained structures we simulate the interaction of a propagating SW with a wrinkle in the CrI$_3$ monolayer. Fig.~\ref{fig_wrinkle}(a) illustrates the considered system. In such a configuration, the position of a generic Cr atom in the curved monolayer can be parametrized as $\textbf{r}'=\textbf{r}+\textbf{u}(\textbf{r})$, where $\textbf{r}=(x, y, 0)$ is the atom position in the pristine (flat) lattice and $\textbf{u}(\textbf{r})=(u_x(\textbf{r}), u_y(\textbf{r}), h(\textbf{r}))$ the position displacement. Here, $u_x$ and $u_y$ represent the in-plane displacement and $h(\textbf{r})$ is a scalar field accounting for out-of-plane deformations. The red and blue shades in Fig.~\ref{fig_wrinkle}~(a) illustrate the expected regions of tensile and compressive strain respectively~\cite{quereda2016strong}. In addition to the local strain, the curvature also induces a rotation in the direction of the magnetic anisotropy, which points orthogonal to the surface of the magnetic film [Fig.~\ref{fig_wrinkle}(a)]. Moreover, due to the non-uniform deformation, the corresponding symmetry-breaking gives rise to localized DMI~\cite{sabani2020releasing}. Figure~\ref{fig_wrinkle}(b) illustrates the expected DMI configuration in the case of a non-uniform uniaxial strain in the wrinkled CrX$_3$ monolayer. DFT calculations show that the DMI increases linearly with local uniaxial strain, with relatively small magnitude of up to $D=0.08$~meV for $6\%$ of local strain~\cite{sabani2020releasing}. 


\begin{figure*}[t!]
\centering
\includegraphics[width=\linewidth]{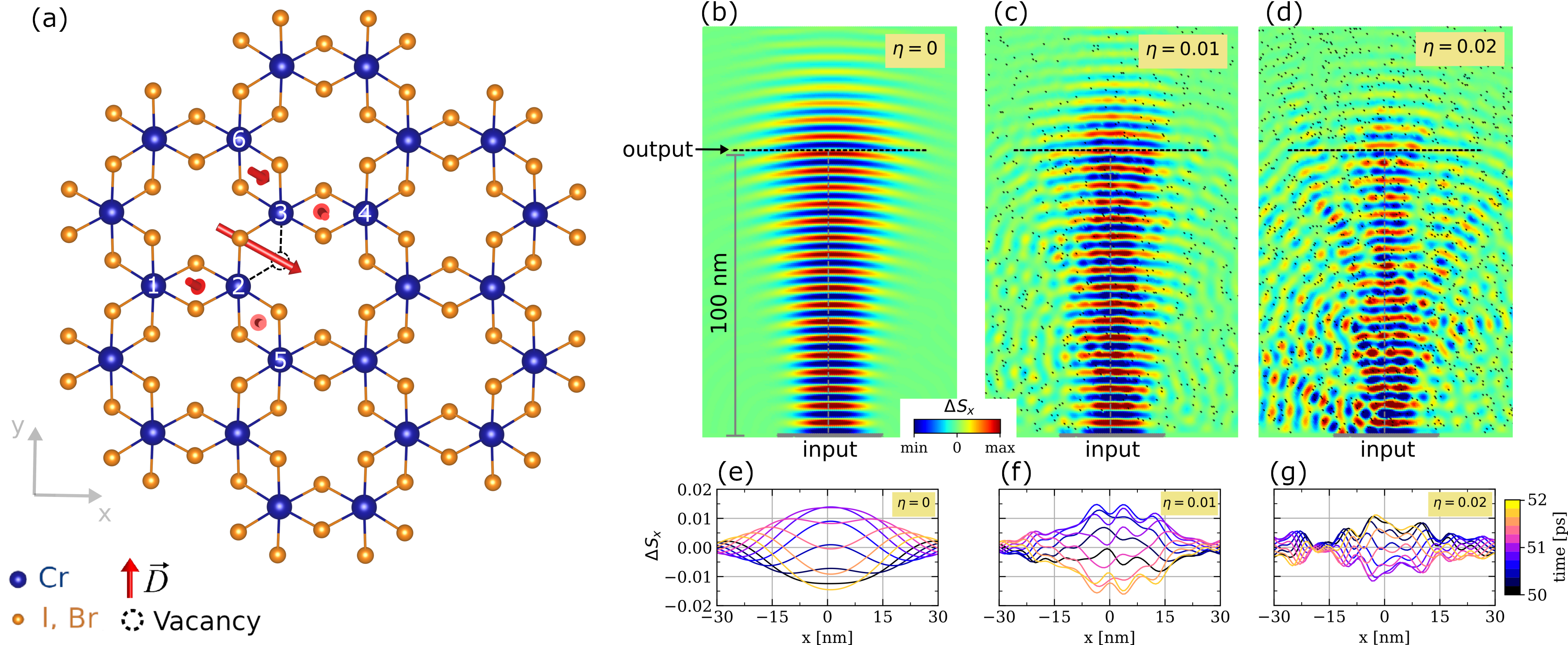}
\caption{ (a) CrX$_3$ lattice in the vicinity of a single halide vacancy. Induced DMI vectors are shown as red arrows. (b-d) Snapshots of simulated SW propagation in presence of randomly distributed defects in CrI$_3$, for defect densities $\eta=0$, $\eta=0.01$, $\eta=0.02$ vacancies per Cr atom, respectively. The input SW frequency is $0.5$~THz. (e-g) Corresponding output SW amplitudes for the cases shown in (b-d), respectively, measured at a distance of $100$~nm from the SW source. Colors in (e-g) correspond to different time instances during the $2$~ps SW period. }
\label{fig_def1}
\end{figure*}

\begin{figure}[b!]
\centering
\includegraphics[width=\linewidth]{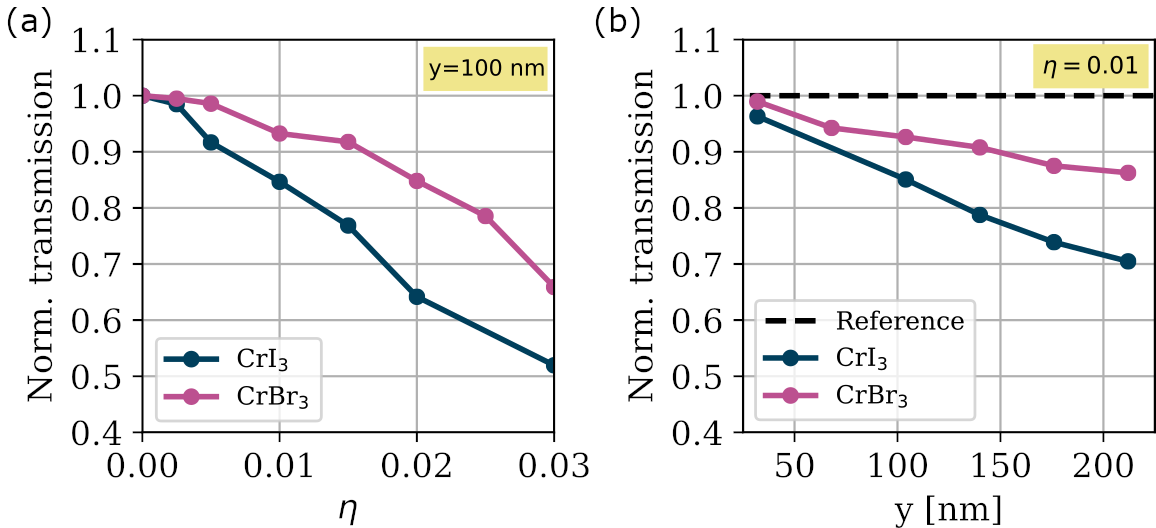}
\caption{(a) The normalized transmission of SWs as a function of the density of defects, $\eta$, measured at a distance of $y=100$~nm from the SW source, with $f=0.5$~THz. (b) SW transmission as a function of the distance to the output, for $\eta=0.01$ and $f=0.5$~THz. Here, the transmission amplitudes are normalized with respect to the case of $\eta=0$, measured at the same distance from the SW source.}
\label{fig_def2}
\end{figure}

\begin{figure*}[t!]
\centering
\includegraphics[width=\linewidth]{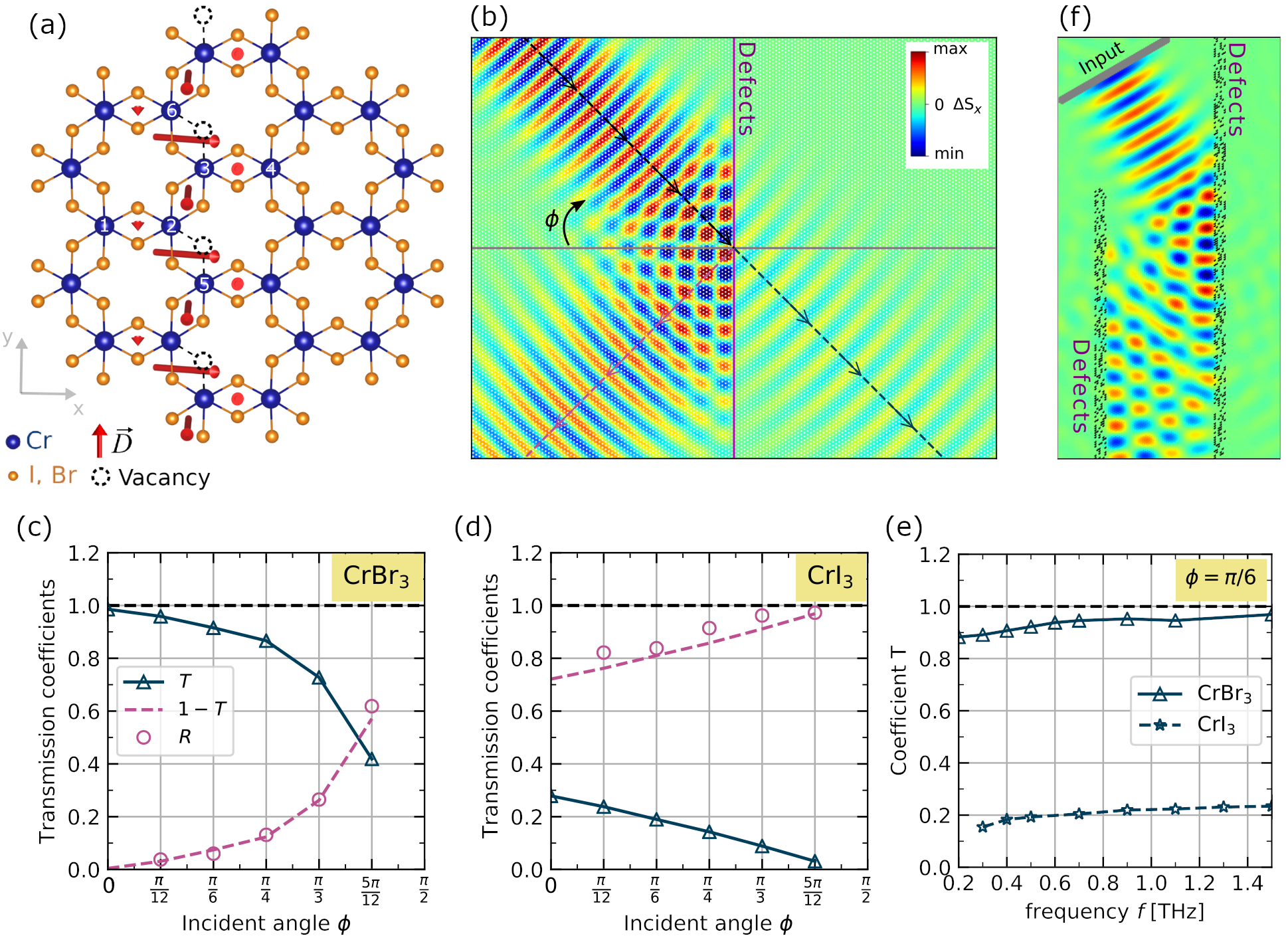}
\caption{(a) CrX$_3$ lattice with a line of halides vacancies. Induced DMI vectors are illustrated as red arrows. (b) Reflection of a SW by a defect line shown in (a). Snapshot captures a spin-wave of frequency $0.5$~THz reaching the defect line in CrI$_3$ under an incident angle of $\phi=\pi/4$. Arrows indicate the SW propagation direction. (c-d) Transmission and reflection coefficients as a function of the incident angle $\phi$ for CrBr$_3$ (c) and CrI$_3$ (d), obtained for SW frequency $f=0.5$~THz and damping $\alpha=0.001$. (e) Frequency dependence of the transmission coefficient $T$ for both CrBr$_3$ and CrI$_3$ at fixed incident angle $\phi=\pi/6$. (f) Snapshot of the simulations for the SW confinement between two realistically broadened defect lines in the magnetic 2D material.}
\label{fig5}
\end{figure*}

First-principles calculations of such a curved structure are very challenging, requiring relaxation of the atomic structure over a large area before calculating locally the magnetic parameters for a number of distinctly different Cr-Cr pairs due to heterostraining. Instead, as a first step in understanding the interaction of SWs in such systems, here we model the material curvature by rotating the local bases of eigenvectors of the magnetic Hamiltonian [Eq.~(\ref{eq.H2})] accordingly to the surface normal, and vary the magnitude of interactions accordingly to the local strain profile of the curvature. Note that the considered magnetic Hamiltonian is completely determined by the exchange and anisotropy matrices and the angle between neighbouring spins, therefore, local deformations can be modeled by correspondingly modifying the interaction matrices and precise atomic positions do not play a role. Fig.~\ref{fig_wrinkle}(c) shows the magnetic ground state obtained in such simulations for a single wrinkle of Gaussian shape $h(\textbf{r}) = h_0e^{-\frac{x^2}{w^2}}$, where $h_0=20$~nm and $w=15$~nm define the height and width of the wrinkle, respectively. The considered strain profile has maximal deformation of $\epsilon=5\%$ on top of the wrinkle~\cite{yang2021strain,quereda2016strong,deng2018strain} and satisfies $\epsilon(x)\propto\partial^2 h(\textbf{r})/\partial x^2$. Notice that the curvature induces local canting in the ground-state magnetization and increases spin-stiffness due to emergent DMI. Fig.~\ref{fig_wrinkle}(d-f) shows snapshots of the simulated propagation of a $0.5$~THz SW (d) across and (e) along the wrinkle, where the SW can either be reflected by or confined along the flexed region, respectively, in comparison with the case of a pristine (flat) sample (f). Recalling the wide range of possibilities for nanoengineered manipulation of 2D materials, this basic example suffices to promote ideas towards flexo-magnonic applications.

\section{. Scattering of spin-waves on lattice defects}\label{sec.defects}

\begin{table}[b!]
\begin{tabular}{lcccccccccc}
\hline\hline
Isolated defect\\
  &  pair     &$D^{x}$& $D^{y}$ & $D^{z}$ & $|D|$   \\
  &  ($i$-$j$)& (meV) & (meV)   & (meV) & (meV)         \\
\hline
CrI$_3$&        &       &       &       & \\
       & (1-2)  &0.29	&-0.20	&-0.53	&0.64 \\
       & (2-3)  &4.05	&-2.33	&-0.39	&4.69  \\
       & (2-5)  &0.17	&0.09	&-0.46	&0.50 \\
       & (3-4)  &0.01	&0.11	&-0.06	&0.12 \\
       & (3-6)  &0.01	&-0.38	&0.19	&0.42 \\
CrBr$_3$&       &       &         &  \\
       & (1-2)  &-0.05	&-0.13	&0.26	&0.29 \\
       & (2-3)  &2.24	&-1.31	&-0.08	&2.60  \\
       & (2-5)  &0.00	&0.06	&-0.04	&0.07 \\
       & (3-4)  &0.02	&0.03	&0.07	&0.08 \\
       & (3-6)  &-0.09	&-0.14	&0.07	&0.18 \\
\hline\hline
Line of defects\\
  &  pair     &$D^{x}$& $D^{y}$ & $D^{z}$ & $|D|$   \\
  &  ($i$-$j$)& (meV) & (meV)   & (meV) & (meV)         \\
\hline
CrI$_3$&        &       &       &       & \\
    & (1-2) & 0.01 & -0.43 & 0.06 & 0.44 \\
    & (2-3) & -0.05 & -0.55 & 2.42 & 2.48 \\
    & (2-5) & 3.61 & -0.27 & 4.23 & 5.57\\
    & (3-4) & -0.09 & 0.03 & -0.23 & 0.25\\
    & (3-6) & 3.61 & -0.27 & 4.23 & 5.57\\
CrBr$_3$&       &       &         &  \\
        & (1-2) & 0.00 & -0.09 & 0.01 & 0.09\\
        & (2-3) & 0.12 & 0.14 & 0.47 & 0.50\\
        & (2-5) & 2.09 & -0.02 & 1.50 & 2.57\\
        & (3-4) & -0.04 & -0.03 & 0.01 & 0.05\\
        & (3-6) & 2.09 & -0.02 & 1.50 & 2.57\\
\hline\hline 
\end{tabular}
\caption{Induced DMI parameters at an isolated defect (halide vacancy) and at a line of such defects in CrI$_3$ and CrBr$_3$, obtained from first-principle calculations \cite{sabani2020releasing}.  The complete exchange matrix for both cases is provided in the supplemental material~\cite{SM}. Pairs (\textit{i-j}) are indicated in Fig.~\ref{fig_def1}~(a) and Fig.~\ref{fig5}~(a). }  
\label{table_defects}
\end{table}

\noindent Vacancies are the most commonly observed defect in 2D materials. They are not only an unwanted product of synthesis or manipulation of the sample, but can also be deliberately artificially induced by vacancy engineering~\cite{jiang2019defect,schuler2019large}. In this section we therefore investigate the interaction of SWs with such defects (halides vacancies) in CrX$_3$ lattice. The local breaking of the inversion symmetry seen by Cr atoms in the vicinity of the lattice defect results in appearance of DMI, as shown in Fig.\ref{fig_def1}(a). It is well known that SWs are strongly affected by the DMI interaction and can even be controllably refracted/reflected at an interface where DMI changes\cite{mulkers2018tunable,wang2018probing}. In the case of lattice defects, \textit{ab initio} calculations reveal DMI vectors of magnitudes up to $D=4.69$~meV for CrI$_3$ and $D=2.60$~meV for CrBr$_3$ in the vicinity of halide vacancies~\cite{sabani2020releasing}. One thus expects pronounced effects of such strong variations of the local magnetic interactions on the SW propagation in the 2D magnetic materials. 

To start with, we consider the scattering of SWs in CrBr$_3$ and CrI$_3$ with different density of randomly distributed halide vacancies $\eta$ (density expressed as the number of vacancies per Cr atom). The used magnetic parameters, as obtained from first-principles calculations, are listed in Table~\ref{table_defects} and in the supplemental material~\cite{SM}. Fig.~\ref{fig_def1}(b-d) shows the snapshots of simulated SW propagation through the CrI$_3$ monolayer for $\eta=0$, $0.01$ and $0.02$, respectively. The corresponding output SW amplitudes, measured at a distance of $100$~nm from the SW source [Fig.\ref{fig_def1}~(e-g)], reveal that the transmitted wave is strongly affected in the case of high density of defects. In order to quantify and compare the SW transmission for different defect densities, we integrate the output SW amplitudes along the direction perpendicular to the SW propagation and normalize it with respect to the case without defects, i.e., for $\eta=0$. Fig.~\ref{fig_def2}~(a) shows such normalized transmission for the $0.5$~THz SWs in CrBr$_3$ and CrI$_3$. Notice that SWs in CrI$_3$ are more affected by the halide vacancies when compared to CrBr$_3$. We relate this property to the large DMI interactions induced in CrI$_3$, which are approximately twice as large as in the CrBr$_3$ sample, predominantly due to stronger spin-orbit coupling on iodine compared to bromine.

Scattering on defects, as shown in Fig.~\ref{fig_def1}, especially over large distances, must clearly be taken into account when designing novel magnonic devices based on 2D materials. Therefore in Fig.~\ref{fig_def2}(b) we show the SW transmission as a function of the output distance, normalized with respect to the case of $\eta=0$, measured at the same distance from the SW source. Notice that the normalized transmission decreases quasi-linearly as a function of the output distance, where the transmission in CrI$_3$ decreases at approximately twice higher rate than in CrBr$_3$, for $\eta=0.01$, thus again reflecting the stronger response of SWs to defects in CrI$_3$. Finally we note to not have observed any significant frequency dependence (for the range of $\SIrange{0.3}{1.5}{}$~THz) in the results presented in this section.

\subsection{Defect-engineering of magnonic circuitry}

\noindent As demonstrated above, the induced strong variations in magnetic parameters around the lattice defects strongly affect the SW propagation. Therefore, a designed pattern of defects, such as a chain of adjacent defects, may be able to e.g. confine the SWs and serve as a waveguide. To illustrate this point, we investigate the interaction of SWs with a row of consecutive halide vacancies in CrX$_3$ lattices, as a possible way of controlling the SW propagation direction in magnetic 2D materials. 

The emergent DMI in the considered structure, as illustrated in Fig.~\ref{fig5}(a), is similar in magnitude to DMI found in the vicinity of isolated defects. The exact values are listed in Table~\ref{table_defects} for comparison, while complete exchange matrices are provided in the supplemental material~\cite{SM}. Fig.~\ref{fig5}(b) shows a snapshot of the simulation for a SW directed across the defect line in CrI$_3$, where we consider a $0.5$~THz SW reaching the defect line under an incident angle of $\phi=\pi/4$. Notice that only a small fraction of the SW is able to cross the defect line while the major part is reflected. In Fig.~\ref{fig5}(c-d) we show the transmission coefficients, defined as~\cite{macke2010transmission,wang2013spin}
\begin{equation}
    T=\left(\frac{\Delta_T}{\Delta_0}\right)^2,
    R=\left(\frac{\Delta_R}{\Delta_0}\right)^2,
\end{equation}
where $\Delta_T$ and $\Delta_R$ are the amplitudes of the transmitted and reflected waves, respectively, and $\Delta_0$ is the corresponding SW amplitude in the absence of defects, calculated at the same distance from the SW source. The reflection and transmission coefficients are defined to satisfy the relation $R=1-T$. In Fig.~\ref{fig5}(c-d) the transmission and reflection coefficients are shown as a function of incident angle $\phi$ for CrBr$_3$ [Fig.~\ref{fig5}(c)] and CrI$_3$ [Fig.~\ref{fig5}(d)]. Notice that SWs experience stronger reflection by the defect line in CrI$_3$ while most part of the SW is transmitted across the vacancy defect line in CrBr$_3$. Once again we relate this property to the larger DMI induced at defects in CrI$_3$. Similar to an interface where DMI changes~\cite{mulkers2018tunable}, the reflection is enhanced with increasing the incident angle from $\phi=0$ towards $\phi=\pi/2$. However, total reflection is not found in our simulations, which is due to a finite cross-talk of spins on two sides of the defect line, where DMI is not present, which is different in the case of a DMI interface. Fig.~\ref{fig5}(e) shows the frequency dependence of the transmission coefficient $T$ for both CrBr$_3$ and CrI$_3$ for a fixed incident angle $\phi=\pi/6$, which shows a weak increase of transmission with increasing SW frequency.

Of course, such a precise pattern of defects is unlikely to be reproduced experimentally by vacancy engineering in a 2D material~\cite{jiang2019defect,schuler2019large}.
\begin{figure*}[t!]
\centering
\includegraphics[width=\linewidth]{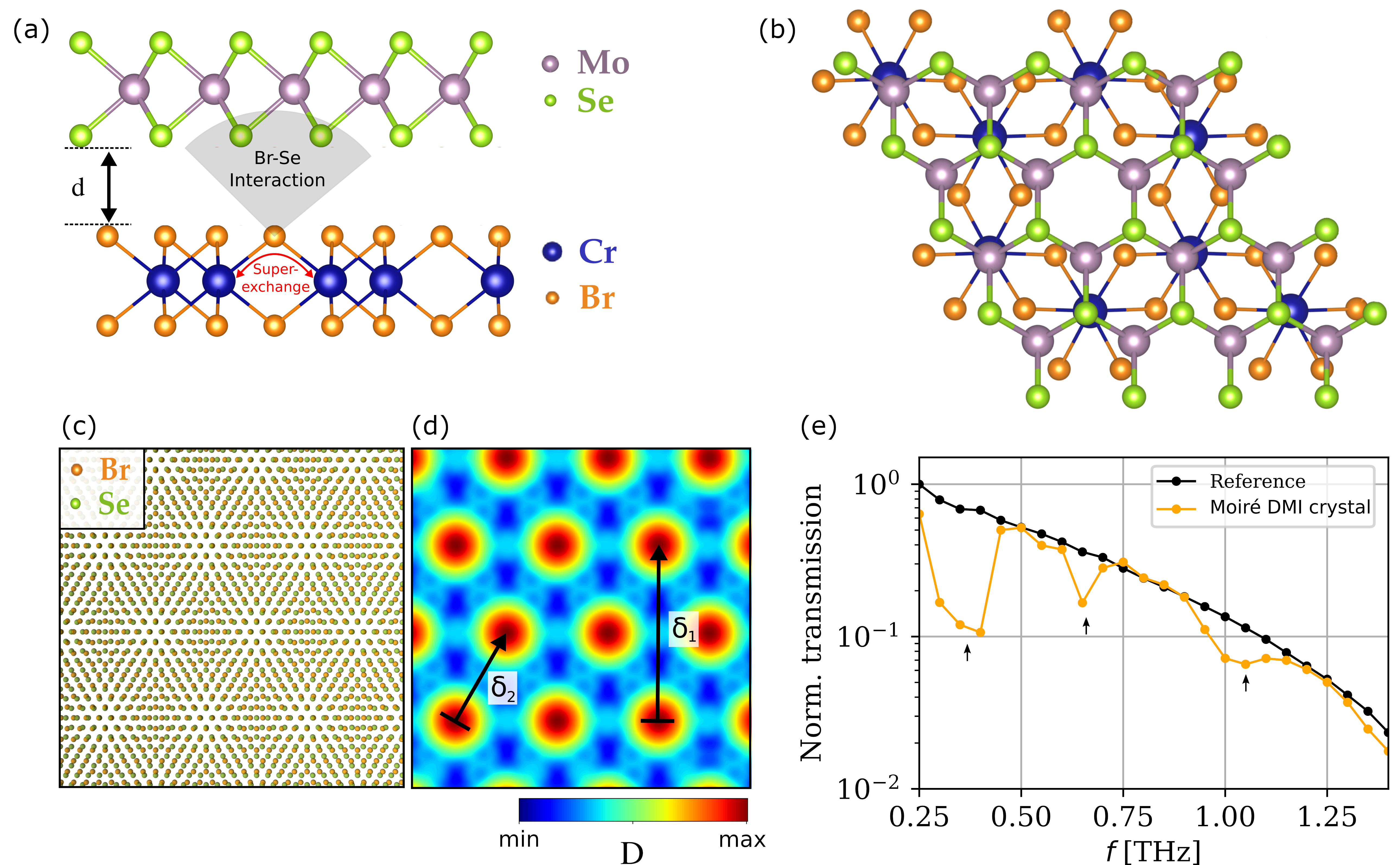}
\caption{Example of a moir\'e magnonic crystal. (a-b) Illustration of a CrBr$_3$/MoSe$_2$ vdW heterobilayer, viewed from the side (a) and the top (b). (c) moir\'e pattern created between adjacent Br and Se sublayers. (d) Corresponding DMI profile emerging in the Cr sublayer from the moir\'e pattern shown in (c). Here, $\delta_1$ and $\delta_2$ denote the two main periodicities of the system. (e) Simulated SW transmission spectra after passing through the moir\'e magnonic crystal (with maximal DMI of 3 meV), in comparison with the reference waveguide (pristine film), where we consider the incident SW propagating along the $\delta_1$ direction. Black arrows indicate the critical frequencies where SWs suffer destructive interference.}
\label{figMagC}
\end{figure*}
However, a broadened defective region can also be made useful in a similar sense, even if induced by e.g. low-energy electron-beam~\cite{jiang2019defect} or by ``scratching'' the material with a scanning tip. To illustrate such a case, in Fig.~\ref{fig5}(f) we show the snapshot of the simulation for the SW propagating in CrI$_3$, across a distribution of defects, randomly placed along 5 nm wide stripes, roughly mimicking surface scratches on the magnetic 2D material. Notice that such a pattern of defects can also confine the SWs, and can be engineered to (de)stimulate self-interference patterns, or control interactions of two incoming spin-waves. The key ingredient for confining and controlling the SWs remains the local variation of the magnetic parameters, which can also be induced by electric gating~\cite{liu2018analysis} or strain-engineering the 2D material, as demonstrated in Sec.~\ref{sec.wrinkle}, albeit on a significantly larger scale than what is offered by defect engineering.  

\section{. Moir\'e van der Waals magnonic crystals}\label{sec.mc}

\begin{figure*}[t!]
\centering
\includegraphics[width=\linewidth]{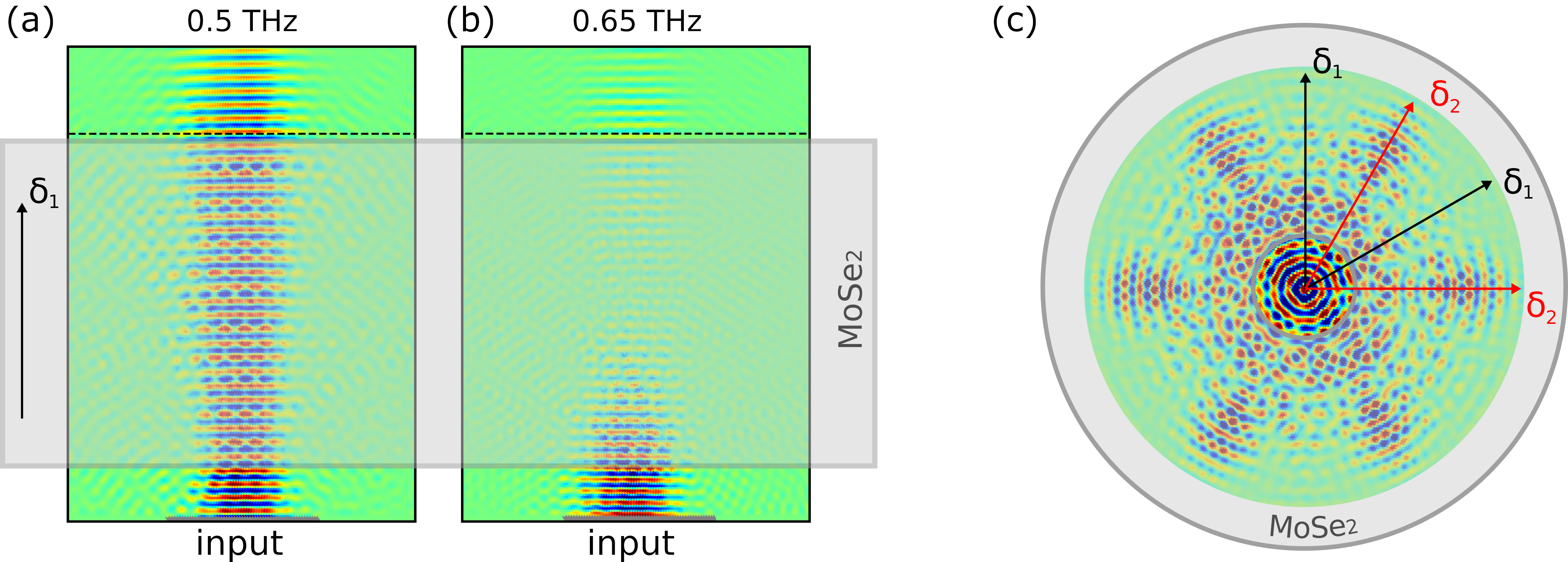}
\caption{ (a-b) Snapshots of SW simulation for the $0.5$ and $0.65$~THz SWs propagating across the moir\'e pattern for the same period of time. The incident SW propagates parallel to the $\delta_1$ direction. (c) Snapshots of the SW simulation for the $0.4$~THz SW propagating across the moir\'e pattern, after being emitted in a radially symmetric fashion from a point source. SW propagation is clearly prohibited along the $\delta_1$ directions, but remains enabled along the $\delta_2$ direction. }
\label{figMagC2}
\end{figure*}

\noindent Magnonic crystals are artificial materials designed in such a way that the magnetic properties of the media are characterized by a periodic lateral variation\cite{chumak2017magnonic}. The SW spectra in such materials exhibit features such as band gaps of wavelengths that are not allowed to propagate, which is useful for magnonic filters as well as information transport and processing based on magnons. Periodic variations of the magnetic parameters in the 2D magnetic materials can be induced in multiple ways, for example, by periodic electric gating~\cite{de2021gate}, strain engineering~\cite{sun2019strain,dai2019strain,peng2020strain}, by growing the 2D material on top of a patterned substrate~\cite{wang2019strain}, or by inducing a checkerboard buckling of the monolayer \cite{mao2020evidence}. However, a magnonic crystal for THz frequencies requires a modulation period of just a few nanometers~\cite{qin2019experimental} [see SW dispersion relation in Fig.~\ref{fig2}], which is beyond the ability of the fabrication methods mentioned above. Instead, we propose here to employ a moir\'e pattern in a van der Waals (vdW) heterobilayer as the source of the periodic modulation in magnetic parameters for the design of magnonic crystals in the high-frequency regime. It is readily established that a moir\'e pattern can induce periodic variation in the magnetic parameters of a 2D material~\cite{xiao2021magnetization,tong2018skyrmions}. Therefore, let us consider the case where the CrX$_3$ monolayer is stacked on top of another, non-magnetic material, in order to produce a moir\'e pattern. For instance, we consider the vdW heterobilayer CrBr$_3$-MoSe$_2$, as illustrated in Fig.~\ref{figMagC}(a,b). The superexchange interaction between Cr atoms, mediated by the Cr-Br-Cr bonds, will be affected by the Br-Se interaction at the interface between monolayer constituents of the heterobilayer, which in turn varies with the local stacking throughout the moir\'e pattern. In other words, the periodicity of the moir\'e structure is directly reflected in a periodic modulation of the resultant magnetic parameters. Fig.~\ref{figMagC}(c) shows the moir\'e pattern created between the Br sublayer (a trigonal lattice with lattice constant $a_\text{Br}=3.698$~\AA) and the Se sublayer (a trigonal lattice with lattice constant $a_\text{Se}=3.289$~\AA)~\cite{wang2016hybrid}. Here the symmetry axes of the two lattices are considered to be aligned, which corresponds to an angle of $30^{\circ}$ between CrBr$_3$ and MoSe$_2$ supercells~\cite{zollner2019proximity}. The mismatch in the lattice constants creates the moir\'e pattern with two characteristic periodicities of $\delta_1\approx5.13$~nm and $\delta_2\approx2.96$~nm, as indicated in Fig.~\ref{figMagC}(d). Such short periodicities found in the moir\'e heterostructures are, therefore, capable of interacting with high frequency SWs and can be adjusted, for example, by fine interlayer twisting, or replacing the MoSe$_2$ layer with a 2D material with a different lattice constant and/or different lattice symmetry.          

Therefore, in the present case the interaction between Br and Se atoms breaks the inversion symmetry seen by the Cr sublayer where magnetism resides, and significantly affects the superexchange owing to the strong spin-orbit coupling in MoSe$_2$~\cite{zhu2011giant,reyes2016spin}. As a result, one can expect a periodic modulation in the DMI parameter throughout the magnetic layer, directly mapped on the moir\'e pattern between Br and Se sublayers. As a first step in understanding the magnonics in such a complex system, and considering that DMI has the main impact on SWs, we will consider a moir\'e periodic modulation only in the DMI parameter, and assume the other magnetic parameters in CrX$_3$ monolayer remain unchanged by vdW heterostructuring. Fig.~\ref{figMagC}(d) shows the used DMI profile, reflecting the moir\'e pattern, according to the expression $D=D_0\exp(-\sqrt{r^2 + d^2}/r_0)$, where $r$ is the lateral distance between Br and Se atoms, $d=3.5$~\AA~is the separation between the layers~\cite{zollner2019proximity} [see Fig.~\ref{figMagC}~(a)] and $D_0$ and $r_0$ are constants that define the DMI magnitude. Here we consider $D_0=3\exp(d/r_0)$~meV and $r_0=0.2$~\AA, such that the maximum magnitude of DMI is $3$~meV. The small value of $r_0$ reflects the short range interaction between neigbouring orbitals, such that the maximal interaction is given when Se atoms are on top of Br ones. Notice that a SW propagating in such a system will experience different periodic modulations, depending on the propagation direction, with respect to the two main periodicities in the system $\delta_1$ and $\delta_2$ [cf. Fig.~\ref{figMagC}(d)]. 

Fig.~\ref{figMagC}(e) shows the simulated SW transmission spectrum of the moir\'e magnonic crystal described above, in comparison with the reference waveguide (pristine CrBr$_3$ monolayer), where we considered the incident SW propagating parallel to the $\delta_1$ direction. Notice that the transmission spectrum of the moir\'e system significantly differs from the pristine medium, as three distinct valleys of prohibited magnon frequencies are visible. Such a spectrum is typical of magnonic crystals\cite{chumak2017magnonic}. Fig.~\ref{figMagC2}(a-b) shows the snapshots of SW simulation for the $0.5$ and $0.65$~THz SWs propagating across the moir\'e DMI pattern, where in the latter case the SWs suffer destructive interference and do not propagate. 


Finally, one notices that the hexagonal symmetry in the modulation of magnetic parameters results in the anisotropic dispersion of the spin waves. Fig.~\ref{figMagC2}(c) shows the snapshots of SW simulation for the $0.4$~THz SW emitted radially across the moir\'e pattern, from a point source. In this case, SW propagation is effectively blocked along the $\delta_1$ directions, as expected from the transmission spectra in Fig.~\ref{figMagC}(e), but remains enabled along the $\delta_2$ directions -- demonstrating the anisotropic SW dispersion characteristic of moir\'e magnonic systems.

\section{. Conclusions}\label{conclusions}

\noindent In summary, we presented an overview of magnonic properties in monolayer chromium trihalides CrBr$_3$ and CrI$_3$, including selected possibilities for tuning the magnonic behavior in these and similar magnetic 2D materials. We revealed that the spin-wave (SW) dispersion relation can be tuned in a broad range of frequencies by straining the 2D magnetic monolayer, and that the SW dispersion has different response to strain in CrBr$_3$ compared to CrI$_3$, which is directly related to differences in their out-of-plane anisotropy. We further demonstrated the possibility of controlling the SW propagation by strain-engineering the 2D material, paving the way towards flexo-magnonic applications. As a next step, we discussed the effect of ever-present structural defects (in this case halide vacancies) on SW propagation in chromium trihalide monolayers. Our calculations show that such lattice defects induce sufficiently large DMI to strongly disturb the SW dynamics, especially in CrI$_3$. However, we also show that a pattern of defects can be designed favorably for magnonic circuitry, as extended defect lines are able to confine SWs and may be useful for SW-guides. Finally, we discuss the spectra of SWs propagating across periodic modulation of the magnetic parameters, induced by a moir\'e pattern in a van der Waals heterobilayer of a magnetic and a non-magnetic 2D material. Such moir\'e heterostructures exhibit the necessary nanoscale modulation period to work as a magnonic crystal for the terahertz SWs characteristic for CrX$_3$ monolayers. Even though we presented only a small fraction of possible manipulations of 2D materials, their pronounced impact on local magnetism and spin excitations, facilitated heterostructuring, as well as the ease of employing additional electronic gating and nanolithography in any desired geometry, promotes magnetic 2D materials to the prime echelon of candidates for terahertz magnonics by design.

\section*{Acknowledgements}
\noindent This work was supported by the Research Foundation - Flanders (FWO-Vlaanderen), the Special Research Funds of the University of Antwerp (BOF-UA), and Brazilian agencies FACEPE, CAPES and CNPq.


\bibliography{references}

\end{document}